\documentclass[symmetry,article,accept,moreauthors,pdftex,10pt,a4paper]{mdpi}

\usepackage{amssymb}
\firstpage{1}
\pubvolume{xx}
\issuenum{1}
\articlenumber{1}
\pubyear{2018}
\copyrightyear{2018}
\externaleditor{Academic Editor: name}
\history{Received: date; Accepted: date; Published: date}

\pdfoutput=1

\newcommand{\kv}{{\mathbf{k}}}
\newcommand{\ii}{{\mathrm{i}}}

\Title{Temperature-dependent $s_\pm \leftrightarrow s_{++}$ transitions in the multiband model for Fe-based superconductors with impurities}

\Author{V.A. Shestakov $^{1}$, M.M. Korshunov $^{1,}$*\orcidA{} and O.V. Dolgov $^{2,3}$}

\AuthorNames{V.A. Shestakov, M.M. Korshunov and O.V. Dolgov}

\address{%
$^{1}$ \quad Kirensky Institute of Physics, Federal Research Center KSC SB RAS, 660036 Krasnoyarsk, Russia\\
$^{2}$ \quad Donostia International Physics Center, 20018 Donostia-San Sebastian, Spain\\
$^{3}$ \quad P.N. Lebedev Physical Institute RAS, 119991 Moscow, Russia}

\corres{Correspondence: mkor@iph.krasn.ru}

\abstract{We study the dependence of the superconducting gaps on both the disorder and the temperature within the two-band model for iron-based materials. In the clean limit, the system is in the $s_\pm$ state with the sign-changing gaps. Scattering by nonmagnetic impurities leads to the change of sign of the smaller gap thus resulting in a transition from the $s_{\pm}$ to the $s_{++}$ state with the sign-preserving gaps. We show here that the transition is temperature-dependent, thus, there is a line of $s_\pm \to s_{++}$ transition in the temperature-disorder phase diagram. There exists a narrow range of impurity scattering rates, where the disorder-induced $s_\pm \to s_{++}$ transition occurs at low temperatures, but then the low-temperature $s_{++}$ state transforms back to the $s_\pm$ state at higher temperatures. With increasing impurity scattering rate, temperature of such $s_{++} \to s_{\pm}$ transition shifts to the critical temperature $T_c$ and only the $s_{++}$ state is left for higher amount of disorder.}


\begin{document}

\section{Introduction}

Iron-based pnictides and chalcogenides attract much attention due to the presence of unconventional superconducting state and peculiar normal state properties \cite{SadovskiPhUsp2008, IzyumovReview2008, IvanovskiiReview2008, PaglioneReview, MazinReview, WenReview, StewartReview, Hirschfeld2011}. Electronic structure of these Fe-based high-$T_c$ superconductors (FeBS) demonstrate several features putting them apart from other conventional and unconventional superconductors. First of all, most of FeBS have similar topology of the Fermi surface that consists of two or three hole pockets centered at $(0,0)$ point and two electron pockets centered at $(\pi,\pi)$ point of the Brillouin zone corresponding to one Fe per unit cell. Exceptions are the cases of the extreme electron or hole doping, where only one group of sheets remains. Second important feature is that all five iron $3d$ orbitals contribute to the formation of the Fermi surface and even a single Fermi pocket is formed by several $d$-orbitals. Thus the system can only be described within a multiorbital and, respectively, multiband model. Minimal model would be a two-band model, such as suggested in Refs.~\cite{Raghu, Efremov2011, KorshunovPhUsp2016}. The third interesting feature is connected with the unconventional superconductivity. While, in general, there are some controversy in  discussion  of the gap symmetry in unconventional superconductors, growing   evidences  in  iron-based  materials  support  sign-changing gap. The structure of the Fermi surface provides practically ideal nesting at the antiferromagnetic wave vector $\mathbf{Q}=(\pi,\pi)$. Long-range spin-density wave antiferromagnetic state is destroyed by doping, however, the enhanced spin fluctuations remain. These fluctuations are claimed to be the main source of the Cooper pairing thus contributing to the formation of the superconducting ground state \cite{Mazin2010, Hirschfeld2011, KorshunovPhUsp2014}. Since spin fluctuations with the large wave vector $\sim\mathbf{Q}$ lead to the repulsive interband Cooper interaction, the gap function have to possess the momentum dependence and change sign between the hole and the electron Fermi surface pockets to compensate the sign of the interaction in the gap equation. The simplest solution is called the $s_\pm$ state and corresponds to the $A_{1g}$ representation with the gap having one sign at hole pockets and opposite sign at electron pockets. At the same time, bands near the Fermi level have a mixed orbital character. Therefore, orbital fluctuations enhanced by, for example, electron-phonon interaction may also lead to a superconductivity \cite{Kontani2010, Onari2012, Yamakawa2017}. These two mechanisms of superconducting pairing differs by the dominating superconducting gap structure: sign-changing $s_{\pm}$ state for the spin fluctuations and sign-preserving $s_{++}$ state for the orbital fluctuations. Alas, superconducting gap structure has not been convincingly determined in experiments yet. Nevertheless, series of experimental observation such as the spin resonance peak in inelastic neutron scattering \cite{Maier2008, Korshunov2008, Christianson2008, Inosov2009}, a quasiparticle interference in tunneling experiments \cite{Wang2009, Gonnelli2009, Szab2009, Zhang2012}, the NMR spin-lattice relaxation rate \cite{Nakai2009, Fukazawa2009}, and the temperature dependence of the penetration-depth \cite{Ghigo2017, GhigoPRB2017, Teknowijoyo2018} are conveniently explained assuming the $s_{\pm}$ state.

Scattering on nonmagnetic impurities has different effect on superconductors with different gap symmetries and structures. The pure attractive interaction, both in the intraband and interband channels, results in the $s_{++}$ state. Its reaction to the disorder is well known since the earlier studies of the conventional $s$-wave superconductivity \cite{Anderson1959, Morosov1979} and the modern treatment of the $s_{++}$ state within the multiband models \cite{Onari2009}. According to the so-called Anderson's theorem \cite{Anderson1959}, in a single-band $s$-wave superconductor, a nonmagnetic disorder does not affect the superconducting critical temperature $T_c$. In the case of unconventional superconductors, Anderson's theorem is violated and the critical temperature is suppressed by a nonmagnetic disorder \cite{Golubov1997} similar to the $T_c$ suppression according to the Abrikosov-Gor'kov theory for magnetic impurities \cite{AbricosovGorkovJETP}. However, series of experiments on FeBS reveals that this suppression is less intensive then predicted by the Abrikosov-Gor'kov theory \cite{Karkin2009, Cheng2010, Li2010, Nakajima2010, Prozorov2014}. One possible reason is that the $s_{\pm}$ state transforms to the $s_{++}$ state at some critical value of the impurity scattering rate $\Gamma^\mathrm{crit}$ \cite{Efremov2011, Yao2012, Chen2013, KorshunovPhUsp2016}. The latter is proportional to the impurity potential and the concentration of impurities $n_\mathrm{imp}$. The transition occurs when one of the gaps, the smaller one, decreases and then changes its sign going through zero. By and large, this happens due to the presence of the subdominant attractive intraband interactions besides the dominant interband repulsion resulting in the $s_\pm$ state in the clean case \cite{KorshunovPhUsp2016}.

In our previous studies \cite{Efremov2011, KorshunovMagn2014, KorshunovPhUsp2016, Shestakov2018}, we have discussed mostly the low temperature behavior of the superconducting gaps. Here we study the effect of nonmagnetic disorder on the two-band $s_\pm$ superconductor for a wide temperature range up to $T_c$. We show that there is a range of $\Gamma$'s near $\Gamma^\mathrm{crit}$, where  although at low temperatures the smaller gap changed its sign ($s_{++}$ state), at higher temperatures the sign is changed back and the system again becomes of $s_\pm$ type staying in this state up to $T_c$.

\section{Model and Method}

Here we use the same two-band model as in Refs.~\cite{Efremov2011, KorshunovPhUsp2016, Shestakov2018} with the following Hamiltonian,

\begin{equation}
H = \sum_{\kv, \alpha, \sigma}\xi_{\kv,\alpha}c_{\kv\alpha\sigma}^{\dagger}c_{\kv\alpha\sigma} + \sum_{\mathbf{R}_i, \sigma, \alpha, \beta}{\mathcal{U}_{\mathbf{R}_i}^{\alpha \beta}c_{\mathbf{R}_i\alpha\sigma}^{\dagger}c_{\mathbf{R}_i\beta\sigma} } + H_{SC}, \label{twoBandHimp}
\end{equation}
where operator $c_{\kv\alpha\sigma}^{\dagger}$ ($c_{\kv\alpha\sigma}$) creates (annihilates) electron with momentum $\kv$, spin $\sigma$, and band index $\alpha = a,b$, $\xi_{\kv,\alpha} = \mathbf{v}_{F \alpha}(\kv - \kv_{F \alpha})$ is the quasiparticles' dispersion linearized near the Fermi level with ${\mathbf{v}}_{F\alpha}$ and ${\kv}_{F\alpha}$ being the Fermi velocity and the Fermi momentum of the band $\alpha$, respectively. Hamiltonian contains the term with the impurity potential $\mathcal{U}$ at site ${\mathbf{R}}_i$ and the term responsible for superconductivity, $H_{SC}$. The exact form of the latter is not important for the present discussion but we assume that it provides superconducting pairing due to the exchange of spin fluctuations (repulsive interaction) and electron-phonon coupling (attractive interaction), see details in Ref.~\cite{KorshunovPhUsp2016}.

To describe the effect of disorder, we use the Eliashberg approach for multiband superconductors \cite{AllenMitrovic}. The connection between the full Green's function $\hat{\mathbf{G}}(\kv, \omega_n)$, the self-energy matrix $\hat{\Sigma}(\kv,\omega_n)$, and the `bare' Green's function,

\begin{equation}
\hat{\mathrm{G}}^{(0)\alpha\beta}(\kv, \omega_n) = \left( i\omega_n\hat{\tau}_0 \otimes \hat{\sigma}_0 - \xi_{\kv \alpha}\hat{\tau}_3 \otimes \hat{\sigma}_0 \right)^{-1} \delta_{\alpha \beta}, \label{G02band}
\end{equation}
is established via the Dyson equation, $\hat{\mathbf{G}}(\kv, \omega_n) = \left( \hat{\mathbf{G}}^{(0)-1}(\kv, \omega_n) - \hat{\mathbf{\Sigma}}(\kv, \omega_n) \right)^{-1}$, where $\omega_n = (2n+1)\pi T$ is Matsubara frequency. Green's function is the matrix in the band space (denoted by bold face) and combined Nambu and spin spaces (denoted by hat). The Pauli matrices $\hat{\tau}_i$ and $\hat{\sigma}_i$ refers to Nambu and spin spaces, respectively.

Assuming the isotropic superconducting gaps at the Fermi surface sheets, we set the impurity self-energy to be independent on the momentum $\kv$ but preserving the dependence on the frequency and the band indices,

\begin{equation}
\hat{\mathbf{\Sigma}}(\omega_n) = \sum_{i = 0}^{3}{ \Sigma_{(i)\alpha \beta}(\omega_n)\hat{\tau}_i }.
\end{equation}
Therefore the task is simplified by averaging over $\kv$. As a result, all equations are written in terms of the quasiclassical $\xi$-integrated Green's functions,

\begin{equation}
\hat{\mathbf{g}}(\omega_n) = \int{ d\xi(\kv) \hat{\mathrm{G}}(\kv, \omega_n) } = \left( \begin{matrix} \hat{\mathrm{g}}_{an} & 0 \\ 0 & \hat{\mathrm{g}}_{bn} \end{matrix} \right), \label{xiIntegrated}
\end{equation}	
where

\begin{equation}
\hat{\mathrm{g}}_{\alpha n} = g_{0\alpha n}\hat{\tau}_0 \otimes \hat{\sigma}_0 + g_{2\alpha n}\hat{\tau}_2 \otimes \hat{\sigma}_2, \label{gAlpha}
\end{equation}
Here $g_{0\alpha n}$ and $g_{2\alpha n}$ are the normal and anomalous (Gor'kov) $\xi$-integrated Green's functions in the Nambu representation,

\begin{align}
&g_{0 \alpha n} = -\frac{ \ii \pi N_{\alpha} \tilde{\omega}_{\alpha n} }{ \sqrt{\tilde{\omega}_{\alpha n}^2 + \tilde{\phi}_{\alpha n}^2} }, &g_{2 \alpha n} = -\frac{ \pi N_{\alpha} \tilde{\phi}_{\alpha n} }{ \sqrt{\tilde{\omega}_{\alpha n}^2 + \tilde{\phi}_{\alpha n}^2} }. \label{g0g2}
\end{align}
They depend on the density of states per spin at the Fermi level of the corresponding band ($N_{a,b}$), and on the order parameter~$\tilde{\phi}_{\alpha n}$ and frequency $\tilde{\omega}_{\alpha n}$. Both the order parameter and the frequency are renormalized by the self-energy,

\begin{align}
&\ii\tilde{\omega}_{\alpha n} = \ii\omega_n - \Sigma_{0 \alpha}(\omega_n) - \Sigma_{0 \alpha}^\mathrm{imp}(\omega_n), \label{tildeOmega} \\
&\tilde{\phi}_{\alpha n} = \Sigma_{2 \alpha}(\omega_n) + \Sigma_{2 \alpha}^\mathrm{imp}(\omega_n), \label{tildePhi}
\end{align}
where $\Sigma_{0(2)\alpha}$ and $\Sigma_{0(2)\alpha}^\mathrm{imp}$ are the parts of the self-energy coming from the pairing interaction (spin fluctuations, electron-phonon coupling, etc.) and impurity scattering, respectively. Impurity scattering will be considered in the $s$-wave channel only. Treating a more complicated impurity potential is a separate cumbersome task, although this goal was pursued by some authors~\cite{Hoyer2015,Scheurer2015,Pogorelov2018}. The order parameter $\tilde{\phi}_{\alpha n}$ is connected with the gap function $\Delta_{\alpha n}$ via the renormalization factor $Z_{\alpha n} = \tilde{\omega}_{\alpha n} / \omega_{n}$, i.e.

\begin{equation}
 \Delta_{\alpha n} = \tilde{\phi}_{\alpha n} / Z_{\alpha n}.
 \label{eq:Delta}
\end{equation}

The part of the self-energy due to the pairing interaction can be written as

\begin{align}
&\Sigma_{0\alpha}(\omega_n) = T\sum_{n',\beta}{ \lambda_{\alpha \beta}^{Z}(n-n') \frac{g_{0\beta n'}}{N_{\beta}} }, \label{Sigma0} \\
&\Sigma_{2\alpha}(\omega_n) = -T\sum_{n',\beta}{ \lambda_{\alpha \beta}^{\phi}(n-n') \frac{g_{2\beta n'}}{N_{\beta}} }, \label{Sigma2}
\end{align}
where $\lambda_{\alpha \beta}^{\phi,Z}(n-n')$ is a coupling function,

\begin{equation}
\lambda_{\alpha \beta}^{\phi,Z}(n-n') = 2\lambda_{\alpha \beta}^{\phi,Z}\int_{0}^{\infty}{ d\Omega \frac{ \Omega B(\Omega) }{ \left( \omega_n - \omega_{n'} \right)^2 + \Omega^2 } },
\end{equation}
that is defined by coupling constants $\lambda_{\alpha \beta}^{\phi,Z}$ and includes density of states $N_{\beta}$ and the normalized bosonic spectral function $B(\Omega)$. The matrix elements $\lambda_{\alpha \beta}^{\phi}$ may be positive (attractive) or negative (repulsive) due to Coulomb repulsion, spin fluctuations and electron-phonon interaction, while matrix elements $\lambda_{\alpha \beta}^{Z}$ are always positive. For simplicity we set $\lambda_{\alpha \beta}^{Z} = \left| \lambda_{\alpha \beta}^{\phi} \right| \equiv \left| \lambda_{\alpha \beta} \right|$.

To calculate the impurity part of the self-energy $\hat{\mathbf{\Sigma}}^\mathrm{imp}$, we use noncrossing diagrammatic approximation that is equivalent to the $\mathcal{T}$-matrix approximation,

\begin{equation}
\hat{\mathbf{\Sigma}}^\mathrm{imp}(\omega_n) = n_\mathrm{imp}\hat{\mathbf{U}} + \hat{\mathbf{U}}\hat{\mathbf{g}}(\omega_n)\hat{\mathbf{\Sigma}}^\mathrm{imp}(\omega_n), \label{Tmatrix}
\end{equation}
where $\hat{\mathbf{U}} = \mathbf{U} \otimes \hat{\tau}_3$ is the matrix of impurity potential, $\left( \mathbf{U} \right)_{\alpha \beta} = \mathcal{U}_{\mathbf{R}_i}^{\alpha \beta}$. Without loss of generality we set $\mathbf{R}_i = 0$. Intra- and interband parts of the impurity potential are given as $v$ and $u$ respectively, $\left( \mathbf{U} \right)_{\alpha \beta} = \left( v - u \right)\delta_{\alpha \beta} + u$. The relation between intra- and interband impurity scattering is set by a parameter $\eta = v / u$.

It is convenient to introduce the generalized cross-section parameter

\begin{equation}
\sigma = \frac{\pi^2 N_a N_b u^2}{1 + \pi^2 N_a N_b u^2} \to \begin{cases} 0, & \mbox{Born limit}, \\ 1, & \mbox{unitary limit} \end{cases}
\end{equation}
and the impurity scattering rate

\begin{equation}
\Gamma_{a(b)} = 2 n_\mathrm{imp} \pi N_{b(a)} u^2 \left(1-\sigma\right) = \frac{2 n_\mathrm{imp} \sigma}{\pi N_{a(b)}} \to  \begin{cases} 2n_\mathrm{imp}\pi N_{b(a)} u^2, \mbox{Born limit}, \\ \dfrac{2n_\mathrm{imp}}{\pi N_{a(b)}}, \mbox{unitary limit}. \end{cases} \label{GammaAB}
\end{equation}

Both are shown to have special values in the limiting cases: (i) the Born limit corresponding to the weak impurity potential ($\pi u N_{a(b)} \ll 1$), and (ii) the unitary limit corresponding to the strong impurity potential ($\pi u N_{a(b)} \gg 1$).

Equations presented above are general and can be applied to a wide range of multiband systems. Below we concentrate on iron-based superconductors within the minimal two-band model. Bands are labelled as $a$ and $b$. Spin fluctuation spectrum enters through the bosonic spectral function $B(\Omega)$ \cite{Parker2008, Popovich2010, Charnukha2011}; the superconducting interaction is also determined by the choice of the coupling constants $\lambda_{\alpha \beta}$.

\section{Results and Discussions}

For the calculations, we choose the ratio between densities of states of different bands as $N_{b}/N_{a} = 2$ and set the values of the coupling constants to be ($\lambda_{aa};\lambda_{ab};\lambda_{ba};\lambda_{bb}$) $=$ ($3;-0.2;-0.1;0.5$), so the averaged coupling constant is positive, $\langle \lambda \rangle = \left( \lambda_{aa} + \lambda_{ab} \right)N_{a}/N + \left( \lambda_{ba} + \lambda_{bb} \right)N_{b}/N > 0$, where $N = N_a+N_b$. This set of parameters leads to the $s_{\pm}$ superconducting state with unequal gaps -- larger positive (band $a$) and smaller negative (band $b$) gap -- and the critical temperature $T_{c0} = 40$ K in the clean limit.

\begin{figure}[H]
\centering
\includegraphics[width=0.7\linewidth]{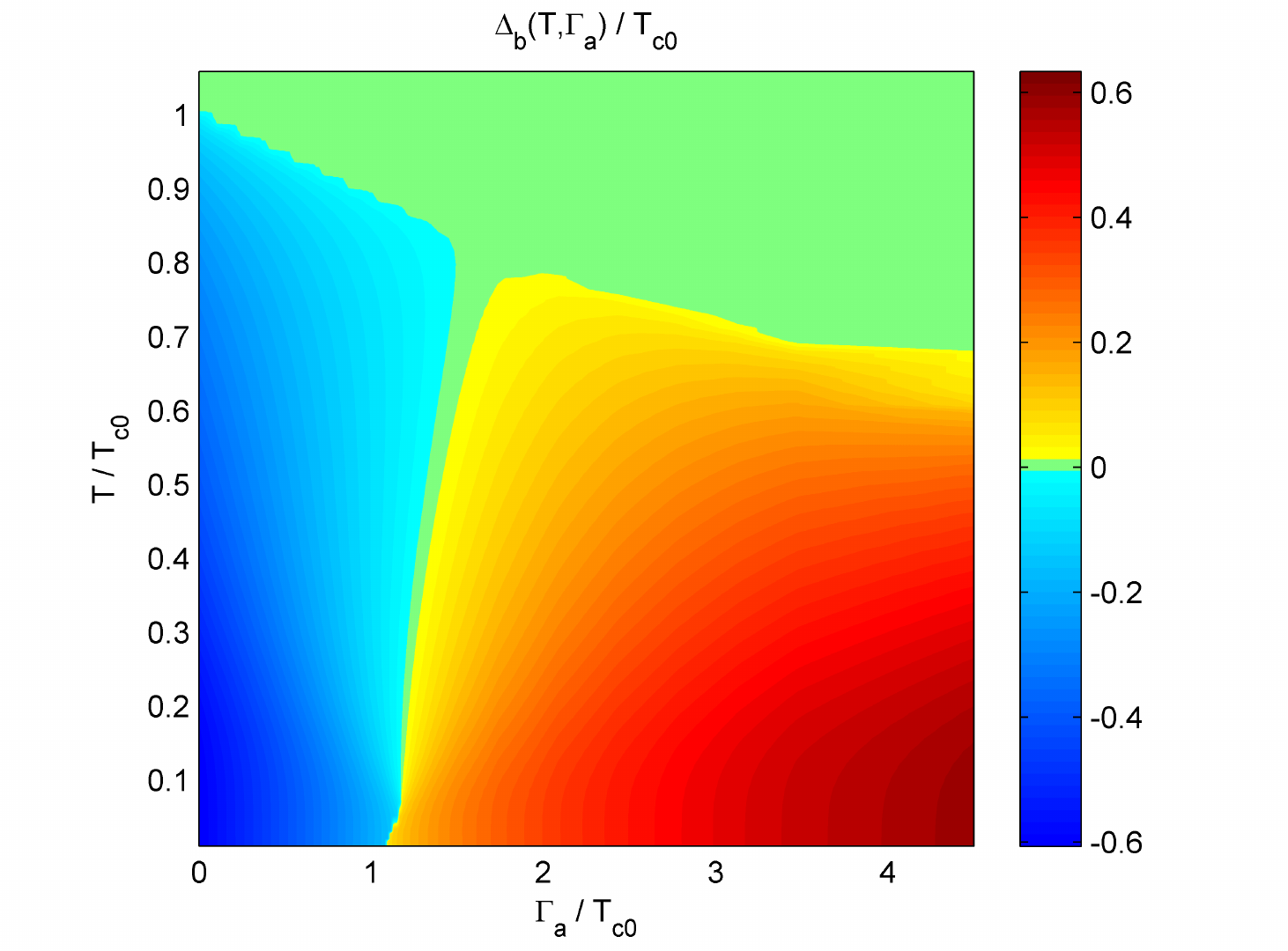}
\caption{Dependence of the lowest-frequency Matsubara gap function $\Delta_{b,n=0}$, indicated by the color code, for the band $b$ on the scattering rate $\Gamma_a$ and the temperature $T$ in the Born limit, $\sigma = 0.0$. All quantities are normalized by $T_{c0}$. Green color marks the state with the vanishingly small gap, $\Delta_{b,n}<10^{-3} T_{c0}$.}
\label{fig:DeltaGammaTsigma0}
\end{figure}

\begin{figure}[H]
\centering
\includegraphics[width=0.7\linewidth]{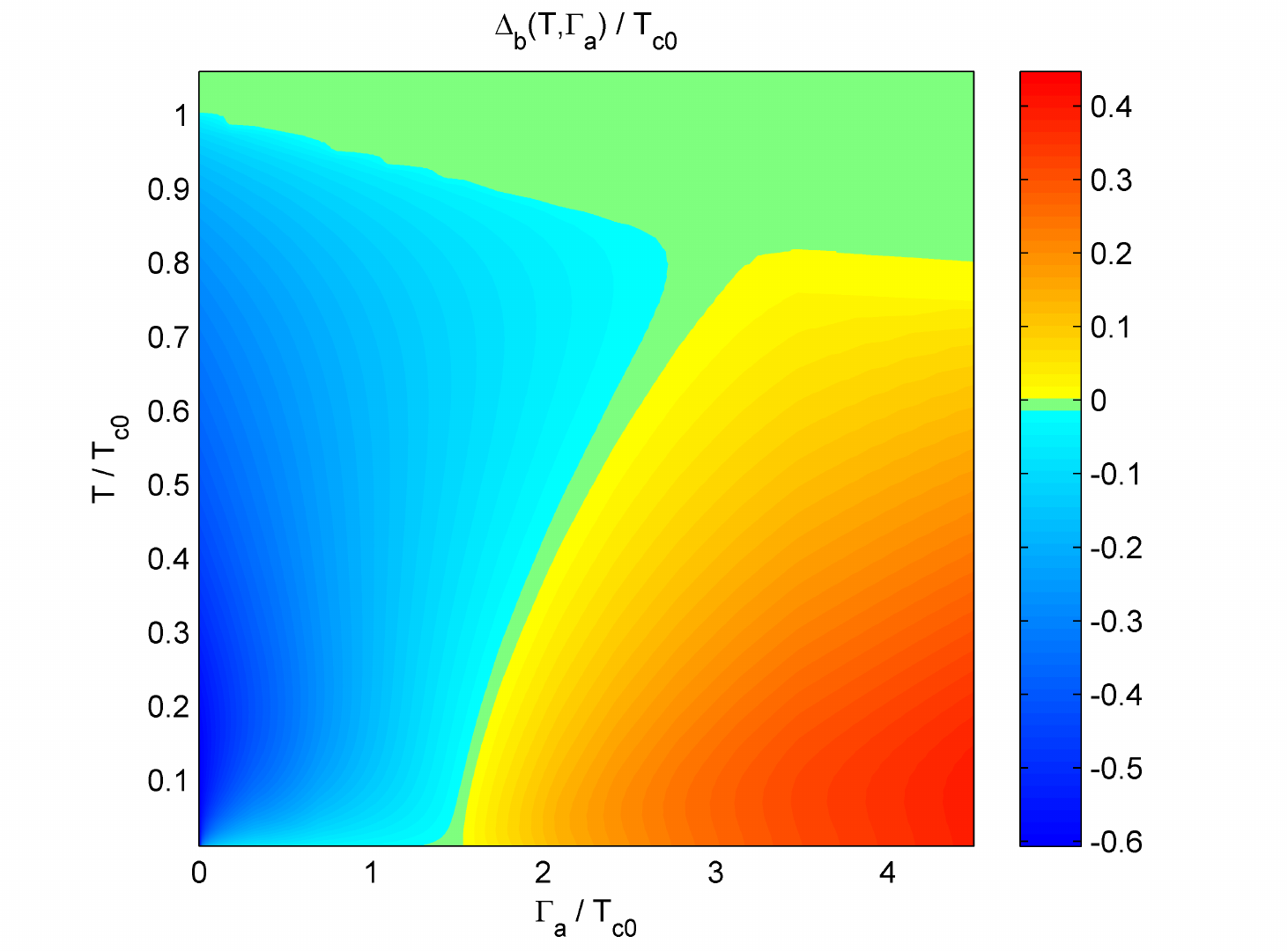}
\caption{Dependence of the lowest-frequency Matsubara gap function $\Delta_{b,n=0}$, indicated by the color code, for the band $b$ on $\Gamma_a$ and $T$ in the intermediate scattering limit, $\sigma = 0.5$. All quantities are normalized by $T_{c0}$. Green color marks the state with the vanishingly small gap, $\Delta_{b,n}<10^{-3} T_{c0}$.}
\label{fig:DeltaGammaTsigma05}
\end{figure}

We present the smaller gap $\Delta_{b,n}$ for the lowest Matsubara frequency, $n=0$, as a function of the scattering rate $\Gamma_a$ and the temperature $T$ in two different cases: (i) weak scattering with $\sigma = 0$ (Born limit), see Figure~\ref{fig:DeltaGammaTsigma0}, and (ii) intermediate scattering limit with $\sigma = 0.5$, see Figure~\ref{fig:DeltaGammaTsigma05}. The larger gap, $\Delta_{a,n}$, is always positive and, therefore, there is a line of $s_\pm \to s_{++}$ transition corresponding to the change of $\Delta_{b,n}$ sign from negative to positive. The transition goes through the gapless state with the finite larger gap and the vanishing smaller gap \cite{Efremov2011}. Note that the line of transition is not vertical in Figures~\ref{fig:DeltaGammaTsigma0} and~\ref{fig:DeltaGammaTsigma05}. Therefore, there is no single critical scattering rate but a temperature-dependent $\Gamma_a^\mathrm{crit}(T)$. Moreover, if we stay at a fixed $\Gamma_a$ in the range $1.1T_{c0} < \Gamma_a < 1.6T_{c0}$ for $\sigma=0$ or $1.5T_{c0} < \Gamma_a < 3.2T_{c0}$ for $\sigma = 0.5$ and increase the temperature, we observe an interesting behavior of the gap. Namely, while at low temperatures the transition to the $s_{++}$ state already took place ($\Delta_{b,n} > 0$), at higher temperatures the system goes back to the $s_\pm$ state ($\Delta_{b,n} < 0$). Thus, there is a temperature-dependent $s_{++} \to s_\pm$ transition. With the increasing $\Gamma_a$, the temperature of this transition is shifted to $T_c$ and the system becomes $s_{++}$ for the whole temperature range.


\begin{figure}[H]
\centering
\includegraphics[width=\linewidth]{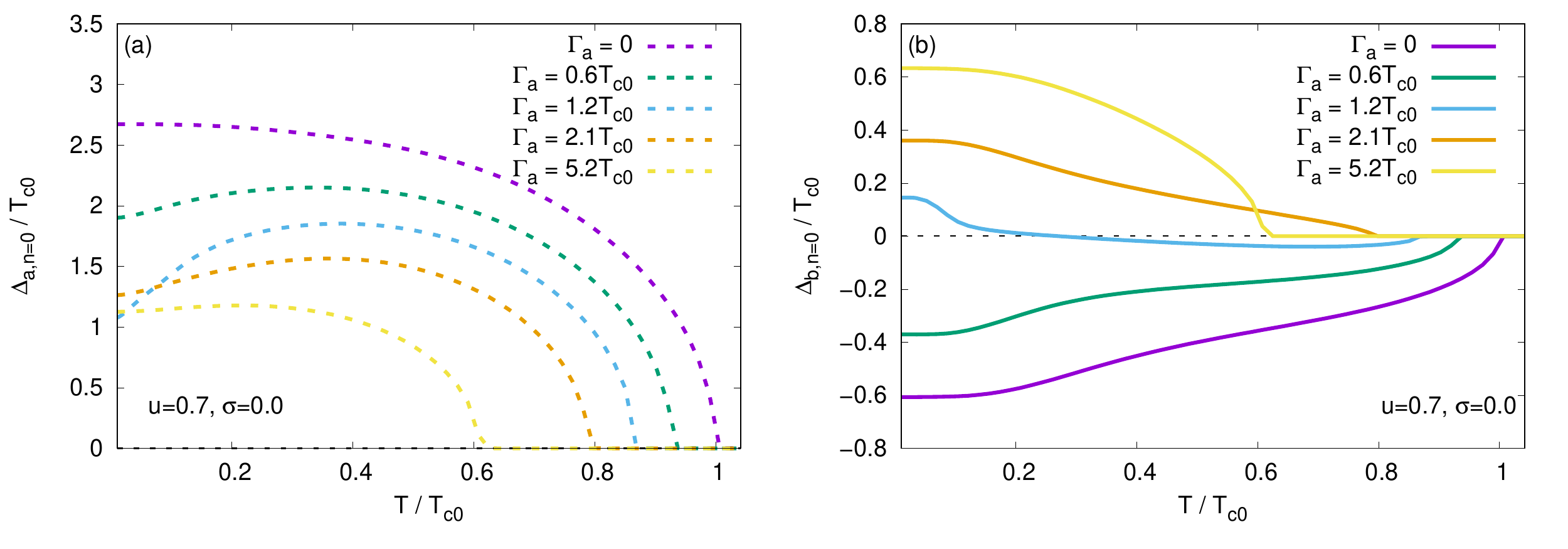}
\caption{Temperature dependence of the lowest-frequency Matsubara gap $\Delta_{\alpha,n=0}$ normalized by $T_{c0}$ for fixed values of $\Gamma_a$ in the Born limit with the band index $\alpha=a$ (\textbf{a}) and $\alpha=b$ (\textbf{b}).}
\label{fig:DeltaTsigma0}
\end{figure}

\begin{figure}[H]
\centering
\includegraphics[width=\linewidth]{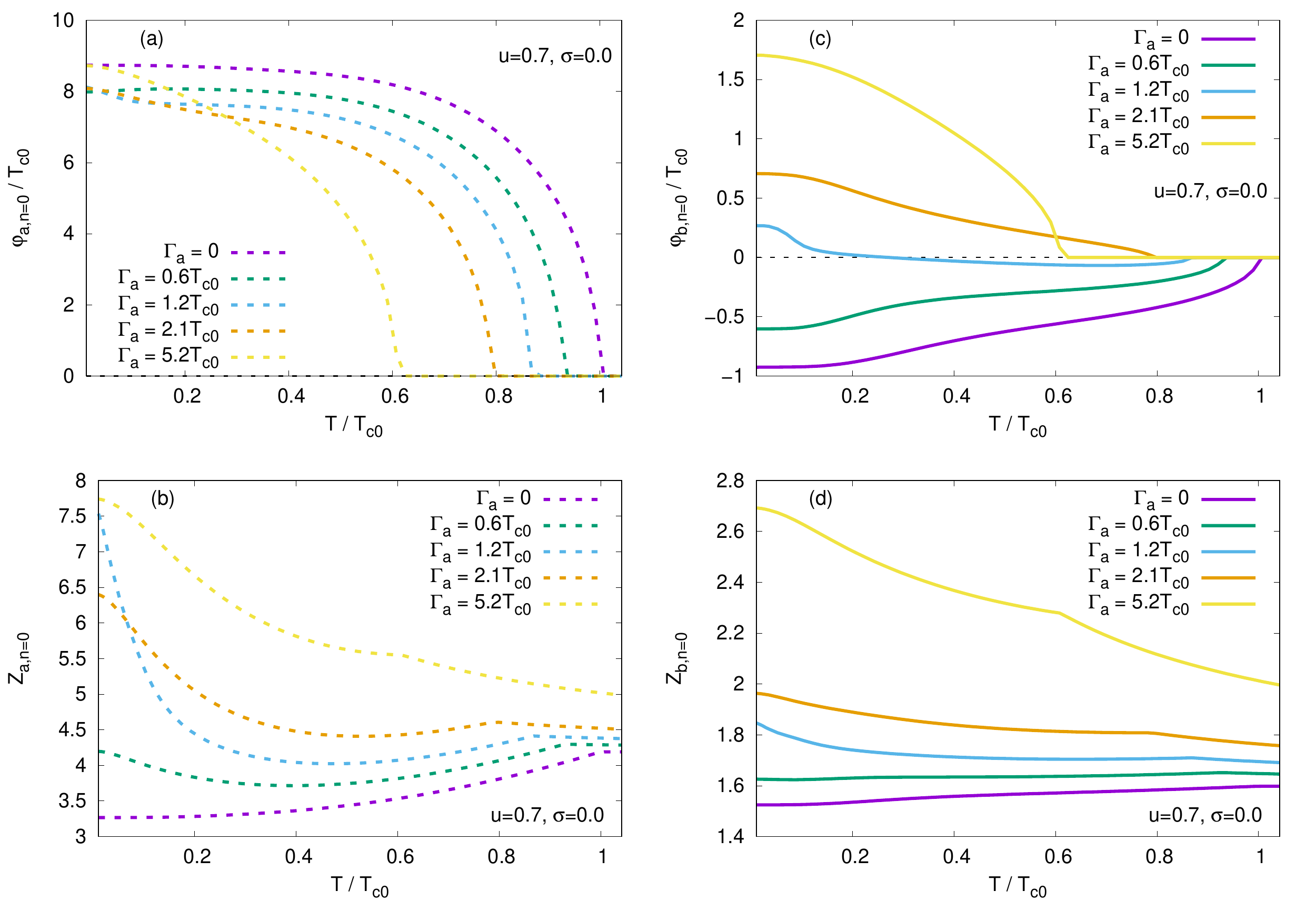}
\caption{Temperature dependence of the lowest-frequency Matsubara order parameter $\tilde\phi_{\alpha,n=0}$ (\textbf{a}),(\textbf{c}) and the renormalization factor $Z_{\alpha,n=0}$ (\textbf{b}),(\textbf{d}), both normalized by $T_{c0}$, for fixed values of $\Gamma_a$ in the Born limit with the band index $\alpha=a$ (\textbf{a}),(\textbf{b}) and $\alpha=b$ (\textbf{c}),(\textbf{d}).}
\label{fig:phiZsigma0}
\end{figure}

\begin{figure}[H]
\centering
\includegraphics[width=\linewidth]{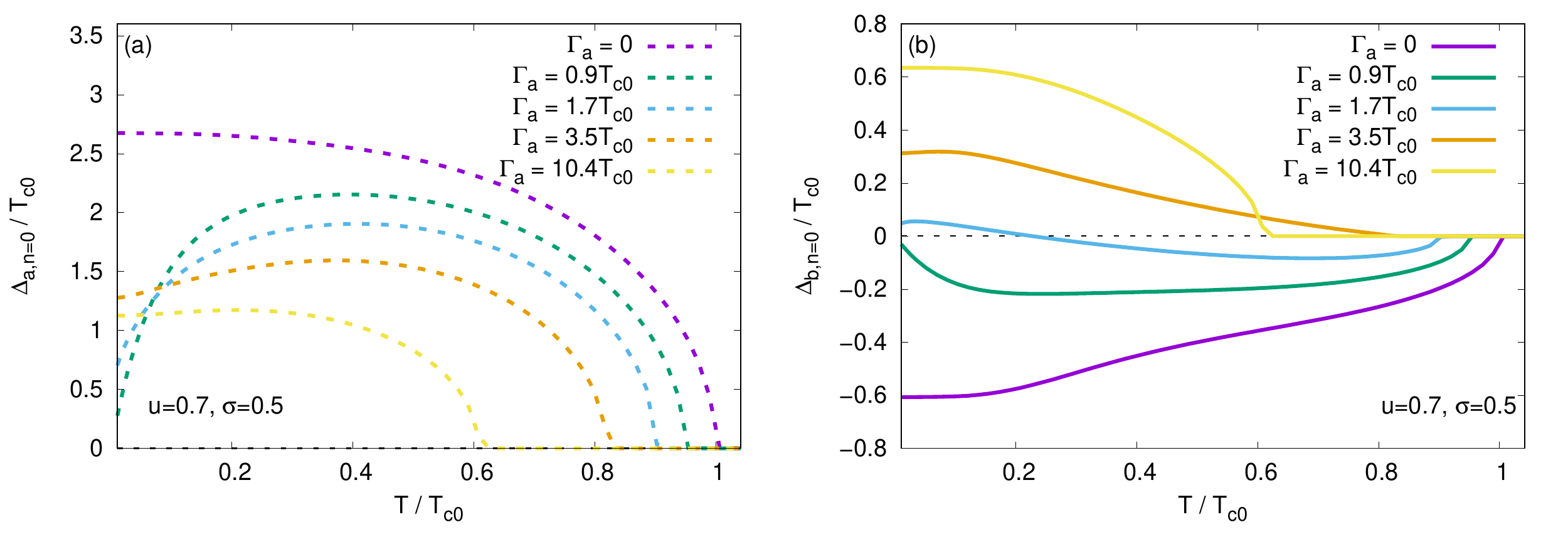}
\caption{Temperature dependence of the lowest-frequency Matsubara gap $\Delta_{\alpha,n=0}$ normalized by $T_{c0}$ for fixed values of $\Gamma_a$ in the intermediate scattering limit ($\sigma=0.5$) with the band index $\alpha=a$ (\textbf{a}) and $\alpha=b$ (\textbf{b}).}
\label{fig:DeltaTsigma05}
\end{figure}

\begin{figure}[H]
\centering
\includegraphics[width=\linewidth]{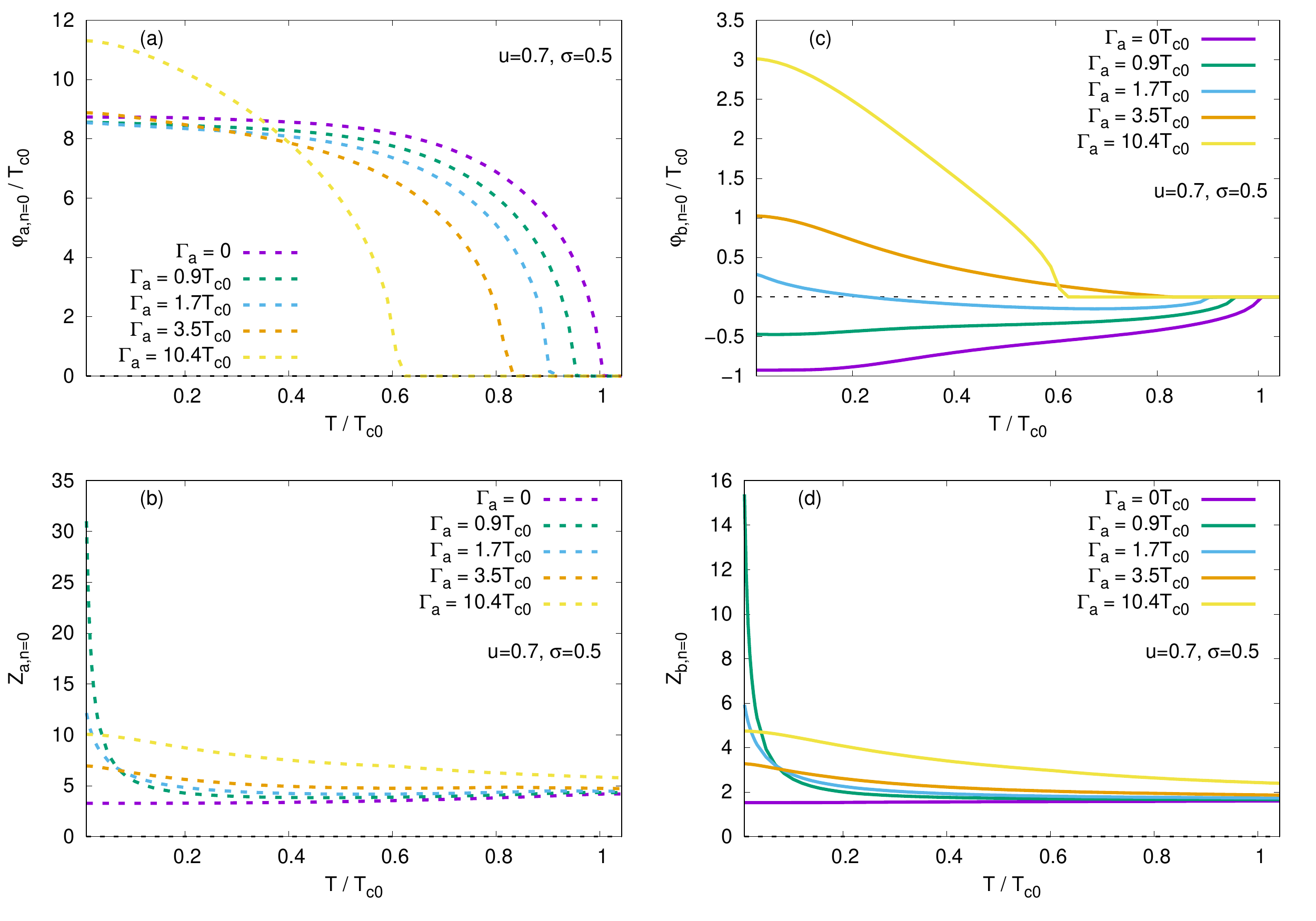}
\caption{Temperature dependence of the lowest-frequency Matsubara order parameter $\tilde\phi_{\alpha,n=0}$ (\textbf{a}),(\textbf{c}) and the renormalization factor $Z_{\alpha,n=0}$ (\textbf{b}),(\textbf{d}), both normalized by $T_{c0}$, for fixed values of $\Gamma_a$ in the intermediate scattering limit ($\sigma=0.5$) with the band index $\alpha=a$ (\textbf{a}),(\textbf{b}) and $\alpha=b$ (\textbf{c}),(\textbf{d}).}
\label{fig:phiZsigma05}
\end{figure}

To illustrate the mentioned points, in Figures~\ref{fig:DeltaTsigma0} and~\ref{fig:phiZsigma0} we present results for the temperature dependence of the gap function $\Delta_{\alpha,n=0}$, order parameter $\tilde{\phi}_{\alpha,n=0}$, and the renormalization factor $Z_{\alpha,n=0}$ for several fixed values of $\Gamma_a$ for $\sigma=0$. Temperature dependencies for $\sigma=0.5$ are shown in Figures~\ref{fig:DeltaTsigma05} and~\ref{fig:phiZsigma05}. The gap in band $a$ has the same positive sign for all values of $\Gamma_a$ and vanishes at $T_c$, see Figures~\ref{fig:DeltaTsigma0}(a) and~\ref{fig:DeltaTsigma05}(a). There is, however, a small range of $\Gamma_a$ values around $\Gamma_a^\mathrm{crit}$, for which the gap is an increasing function of $T$ at low temperatures. At the same time, the order parameter behaves quite conventionally and decreases towards $T_c$, see Figures~\ref{fig:phiZsigma0}(a) and~\ref{fig:phiZsigma05}(a). The unusual temperature dependence of the gap at the lowest Matsubara frequency is due to the renormalization factor $Z_{\alpha,n=0}$ that is shown in Figures~\ref{fig:phiZsigma0}(b) and~\ref{fig:phiZsigma05}(b). That is, according to Equation~(\ref{eq:Delta}), large values of $Z_{\alpha,n}$ taking place at low temperatures lead to the decrease of $\Delta_{\alpha,n}$. For the band $b$ the same effect is not seen in the Born limit, but becomes pronounced in the intermediate scattering limit; compare the gap function in Figure~\ref{fig:DeltaTsigma05}(b) and the order parameter in Figure~\ref{fig:phiZsigma05}(c), and notice also the low-temperature behavior of $Z_{\alpha,n}$ in Figure~\ref{fig:phiZsigma05}(d). Similar behavior of gaps, in principle, may occur from the functional form of gaps at real frequencies, see Eq.~(7) in Ref. \cite{Mikhailovsky1991}.

\begin{figure}[H]
\centering
\includegraphics[width=\linewidth]{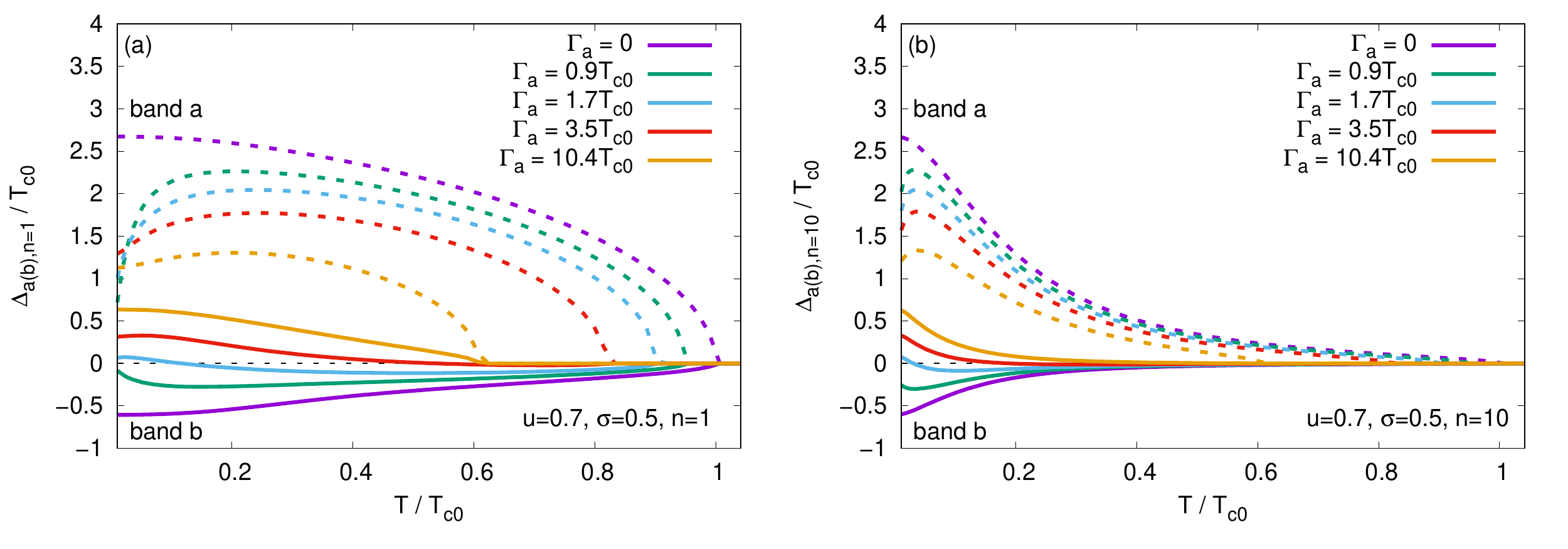}
\caption{Temperature dependence of the gap for higher Matsubara frequencies, $\Delta_{\alpha,n=1}$ (\textbf{a}) and $\Delta_{\alpha,n=10}$ (\textbf{b}), normalized by $T_{c0}$ for fixed values of $\Gamma_a$ in the intermediate scattering limit ($\sigma=0.5$). Gaps corresponding to the band index $a$ (band index $b$) are shown by dashed (solid) curves.}
\label{fig:DeltaTsigma05n1n10}
\end{figure}

In the clean limit and for the small $\Gamma_a$, the sign of the smaller gap $\Delta_{b,n}$ is negative at all temperatures, see Figures~\ref{fig:DeltaTsigma0}(b) and~\ref{fig:DeltaTsigma05}(b). With the increase of the impurity scattering rate, the gap at low temperatures changes sign while at higher temperatures the sign is either reversed again (small $\Gamma_a$) or the gap vanishes ($\Gamma_a \gtrsim 2 T_{c0}$). Therefore, the transition from the $s_\pm$ state to the $s_{++}$ state is characterized by two parameters, namely, the critical scattering rate $\Gamma_a^\mathrm{crit}$ and the critical temperature $T^\mathrm{crit}$. The latter changes from zero to $T_c$. Thus the $s_{++}$ state becomes dominant in the initially clean $s_\pm$ system for $\Gamma_a > \Gamma_a^\mathrm{crit}$ and $T < T^\mathrm{crit}$. This is true in both Born limit and the intermediate scattering limit. Also, we have checked that the similar behavior holds for the higher Matsubara frequencies, see Figure~\ref{fig:DeltaTsigma05n1n10} for the illustration of the gaps behavior for $n=1$ and $n=10$.

\begin{figure}[H]
\centering
\includegraphics[width=0.6\linewidth]{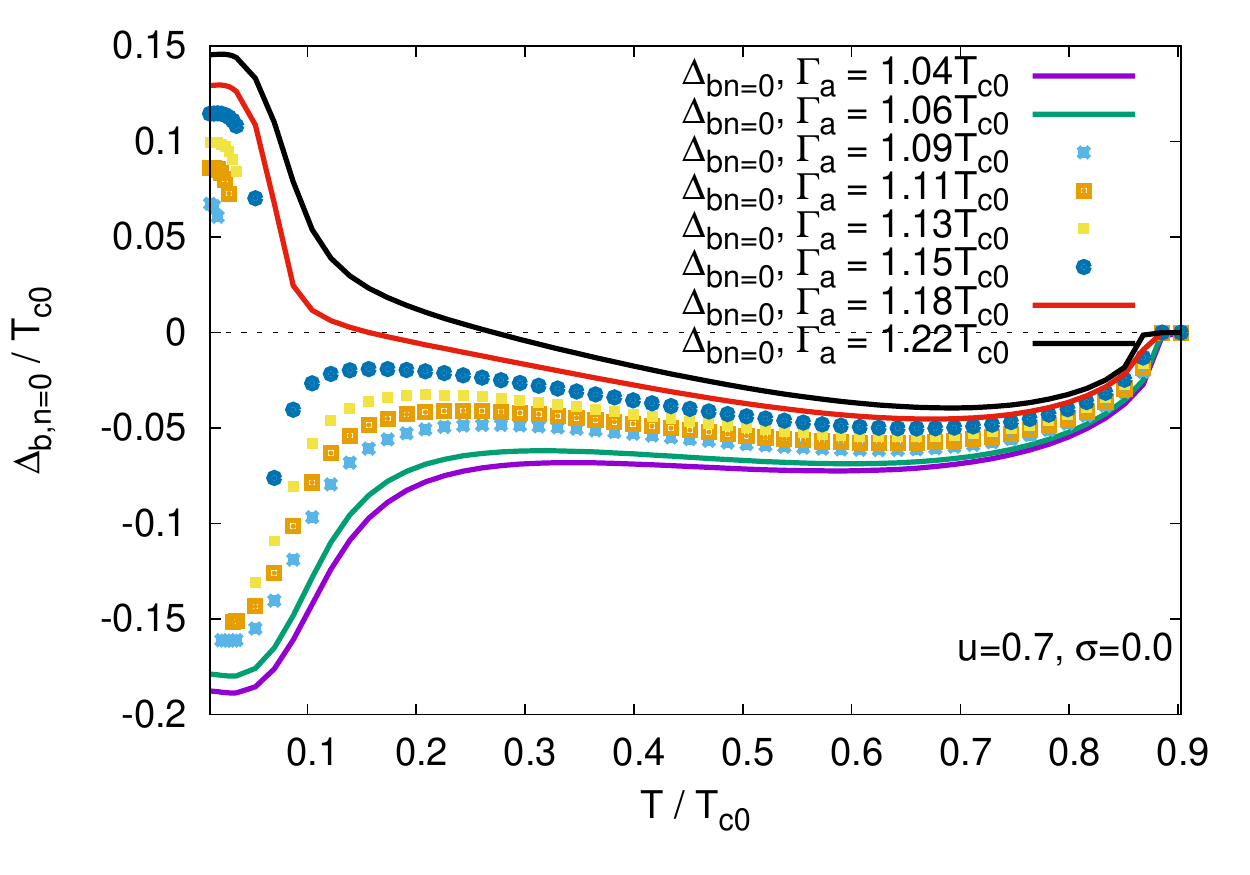}
\caption{Temperature dependencies of lowest-frequency Matsubara gap $\Delta_{b,n=0}$ normalized by $T_{c0}$ in the Born limit for several values of $\Gamma_a$. Solid curves correspond to a smooth evolution of the gap in the $s_\pm$ state and across the $s_{++} \to s_\pm$ transition while the temperature dependencies with the discontinuous jump of the gap are shown by symbols.}
\label{fig:DeltaTsigma0jump}
\end{figure}

Previously we have found a steep change in the smaller gap as a function of the scattering rate in the weak scattering limit \cite{Shestakov2018}. Here we observe that the same jump is present in the temperature dependence of the gap, see Figure~\ref{fig:DeltaTsigma0jump}, where $\Delta_{b,n}$ in the Born limit is shown. Discontinuous jump in the temperature dependence of the smaller gap appears at $T<0.1T_{c0}$.

\section{Conclusions}

We considered the two-band model for FeBS that has the $s_\pm$ superconducting ground state in the clean limit. Here we studied dependence of the superconducting gaps $\Delta_{\alpha,n}$ on both the temperature and the nonmagnetic impurity scattering rate. We show that the disorder-induced transition from $s_\pm$ to $s_{++}$ state is temperature-dependent. That is, in a narrow region of scattering rates, while the ground state is $s_{++}$, it transforms back to the $s_\pm$ state at higher temperatures up to $T_c$. With the increasing impurity scattering rate, temperature of such a $s_{++} \to s_{\pm}$ transition shifts to the critical temperature $T_{c}$. The $s_\pm \to s_{++}$ transition is characterized by two parameters: (i) the critical scattering rate $\Gamma_a^\mathrm{crit}$ and (ii) the critical temperature $T^\mathrm{crit} \leq T_c$. The $s_{++}$ state becomes dominant in the initially clean $s_\pm$ system for $\Gamma_a > \Gamma_a^\mathrm{crit}$ and $T < T^\mathrm{crit}$.
The similar situation takes place for the case of the magnetic disorder where the $s_\pm$ state at low temperatures occurring due to the $s_{++} \to s_\pm$ transition \cite{KorshunovMagn2014}, at higher temperatures may transform back to the $s_{++}$ state.

Experimentally, one can observe the reentrant $s_{\pm}$ state by increasing the temperature for the fixed amount of disorder that results in the low-temperature $s_{++}$ state. For example, the spin resonance peak in the inelastic neutron scattering should be absent in the low-temperature $s_{++}$ state, but have to appear in the $s_{\pm}$ state at higher temperatures \cite{Maier2008, Korshunov2008, Hirschfeld2011, KorshunovPhUsp2014}. Temperature dependence of the penetration depth should also bear specific signatures of the gapless behavior accompanying the $s_{++} \to s_\pm$ transition \cite{KorshunovPhUsp2016}.

\vspace{6pt}



\authorcontributions{All author contributed equally.}

\funding{This work was supported in part by the Russian Foundation for Basic Research (grant 16-02-00098), Presidium of RAS Program for the Fundamental Studies \#12, and ``BASIS'' Foundation for Development of Theoretical Physics and Mathematics. MMK acknowledges support by the Gosbudget program \# 0356-2017-0030.}


\conflictsofinterest{The authors declare no conflict of interest.}

\abbreviations{The following abbreviations are used in this manuscript:\\

\noindent
\begin{tabular}{@{}ll}
FeBS & Fe-based superconductors \\
NMR & Nuclear Magnetic Resonance
\end{tabular}}


\reftitle{References}


\externalbibliography{yes}
\bibliography{References}


\end{document}